\documentclass[aps, twocolumn,
superscriptaddress,showpacs,showkeys,amsmath,amssymb,floatfix]{revtex4}
\usepackage{color}
\usepackage{graphicx}
\usepackage{epsfig}
\usepackage{dcolumn}
\usepackage{bm}
\usepackage{mathtools}
\usepackage{amssymb}
\usepackage{calrsfs}
\DeclareMathAlphabet{\pazocal}{OMS}{zplm}{m}{n} 
\usepackage{amsmath}
\usepackage{subfigure}
\newcommand{\be}{\begin{equation}}
\newcommand{\ee}{\end{equation}}
\newcommand{\bea}{\begin{eqnarray}}
\newcommand{\eea}{\end{eqnarray}}

\newcommand{\tK}{{\tilde K}}

\newcommand{\tF}{{\tilde F}}

\newcommand{\tM}{{\tilde M}}

\newcommand{\pro}{\partial}

\newcommand{\ba}{\begin{array}}
\newcommand{\ea}{\end{array}}

\newcommand{\nn}{\nonumber}

\newcommand{\Ka}{\mathcal{K}}

\newcommand{\Da}{\mathcal{D}}

\newcommand{\Fa}{\mathcal{F}}

\begin{document}
\title{Microscopic vacuum structure in a pure QCD}
\author{D.G. Pak}
\affiliation{Institute of Modern Physics, Chinese Academy of Sciences,
Lanzhou 730000, China}
\affiliation{Chern Institute of Mathematics, Nankai University, Tianjin 300071, China}
\author{P.M. Zhang}   
\affiliation{Institute of Modern Physics, Chinese Academy of Sciences,
Lanzhou 730000, China}
\affiliation{University of Chinese Academy of Sciences, Beijing 100049, China}
\begin{abstract}
       We propose a class of stationary color magnetic solutions in a pure quantum chromodynamics (QCD) 
       which are stable under gluon quantum fluctuations. This resolves a long-standing problem of microscopic 
       description of a stable non-trivial vacuum in a pure QCD. The solutions represent fixed points in a full 
       space of gauge fields under Weyl group transformations. An important feature of the solutions is the 
       phenomenon of Weyl invariant Abelian projection and Abelian dominance in the vacuum structure. 
       As an application of our approach we consider an effective Lagrangian of pure Abelian glueballs in the 
       presence of vacuum gluon condensate.
       \end{abstract}
\maketitle
      There is a common belief that the origin of quark and color confinement lies in the 
   vacuum structure which possesses a number of important features.  
   First, the vacuum is formed by means of condensation
   of some topological defects: monopoles \cite{nambu74,mandelstam76,polyakov77,thooft81}, 
   vortices \cite{niel-oles, amb-oles2,chernodub14}, center vortices
    \cite{centervort1,centervort2,centervort3}, dyons  \cite{diak-petrov} etc. 
    A key long-standing problem in such vacuum formation is the quantum instability
    of the vacuum condensate \cite{savv,N-O}. Another important 
   property is an intimate relationship between the gauge invariance of the vacuum 
   and color confinement \cite{polyakov77}.
   At last, the vacuum structure in the confinement phase should admit the `t Hooft Abelian projection \cite{thooft81}. 
   In this Letter we describe microscopic vacuum structure in a pure QCD
   which provides the main attributes of a true vacuum.
   
      In perturbative QCD the one particle quantum states (color gluons) 
   do not represent physical observables. 
   To find classical solutions which produce possible physical quantum
   states after quantization, we follow a guiding principle formulated by H. Weyl:
    {\it ``Whenever you have to do with a structure endowed
entity $\Sigma$ try to determine its group of automorphisms... .
After that, 
investigate symmetric configurations of elements which are 
invariant under a certain subgroup''} \cite{Weyl1952}.
In a case of the group $G=SU(3)$ with a maximal Abelian subgroup $H=U_3(1)\times U_8(1)$
we look for solutions which are invariant under the Weyl group of outer automorphisms 
$\{g \in G/H \}$.
                                                    
   We start with a standard $SU(3)$ Yang-Mills type Lagrangian of a pure QCD  $(\mu,\nu=0,1,2,3; a=1,...8)$
   \bea
&&{\cal L}_{YM} = -\dfrac{1}{4} F_{\mu\nu}^a F^{a\mu\nu}, \label{Lagr0}
\eea
   A generalized axially symmetric Dashen-Hasslacher-Neveu (DHN) 
   ansatz \cite{DHN,plb2018} for non-vanishing components of the 
   gauge potential $A_\mu^a(r,\theta,t)$ in the holonomic basis reads 
%March9SU3WeylSym
\bea
A_r^2&=&K_1,~~A_\theta^2=K_2,~~A_\varphi^1=K_4,~~A_t^2=K_5, \nn \\
A_r^5&=&Q_1,~~A_\theta^5=Q_2,~~A_\varphi^4=Q_4,~~A_t^5=Q_5, \nn \\
A_r^7&=&S_1,~~~A_\theta^7=S_2,~~A_\varphi^6=S_4,~~~A_t^7=S_5, \nn \\
 &&~~A_\varphi^3=K_3,~~~A_\varphi^8=K_8,   \label{SU3DHN}
\eea
where three sets of gauge fields $K_i,Q_i,S_i$ $(i=1,2,4,5)$ 
with proper combinations of the  Abelian fields $K_{3,8}$ form
the DHN ansatz for each of three $I,U,V$ type subgroups $SU(2)$ \cite{neeman99}. 
To implement the Weyl symmetry we impose additional constraints
\bea
&& Q_{1,2,5}=-S_{1,2,5}=-K_{1,2,5},  ~~~K_{3,8}= -\frac{\sqrt 3}{2} K_4, \nn \\
&& Q_{4}= \big (-\frac{1}{2}+\frac{\sqrt 3}{2}\big ) K_4, ~~~ S_{4}= \big (-\frac{1}{2}-\frac{\sqrt 3}{2}\big ) 
K_4.  \label{reduction1}
 \eea
The full ansatz (\ref{SU3DHN}, \ref{reduction1}) is consistent with the original Yang-Mills equations of motion
and leads to four second order differential equations and one constraint \cite{SM2} which admit 
a wide class of regular stationary solutions. A special non-Abelian solution with a finite energy density
has been found  recently in \cite{plb2018,SM2}. We describe a general class of regular solutions 
of non-Abelian and Abelian type with a definite parity under the space reflection $\theta \rightarrow \pi-\theta$. 
In the leading order of Fourier series decomposition the solutions are well approximated 
by the following factorized form with the numeric accuracy 1.5\% 
(the source of such a high accuracy is explained in \cite{plb2018})
\bea 
K_{1,2,4}(r,\theta,t)&=&\tilde K_{1,2,4}( M r,\theta)  \cos (M t), \nn \\
K_5(r,\theta,t)&=&\tilde K_{5}(Mr,\theta)  \sin(M t),  \label{seriesdec}
\eea
where $M$ is a conformal mass scale parameter.
A new solution with the lowest non-trivial polar angle modes is shown in Fig. 1a-d
(numeric details are given in \cite{SM2}).
The solution differs from the one obtianed in \cite{plb2018} by an opposite parity
of the fields $K_{1,2,5}$, 
\begin{figure}[h!]
%Aug9SU3ABCreg-L6C-r05C.mph
\centering
\subfigure[~]{\includegraphics[width=42mm,height=30mm]{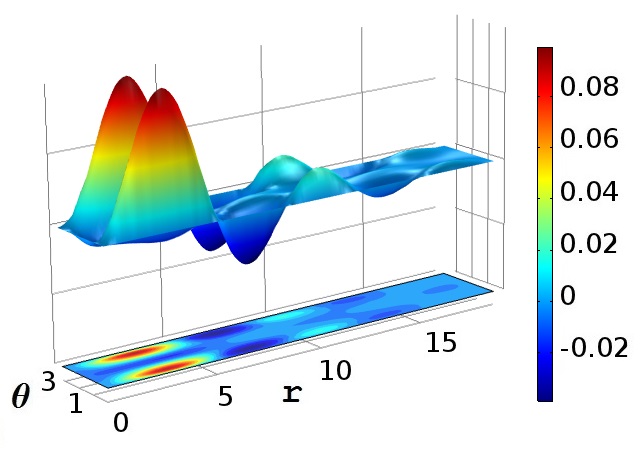}}
\hfill
\subfigure[~]{\includegraphics[width=42mm,height=30mm]{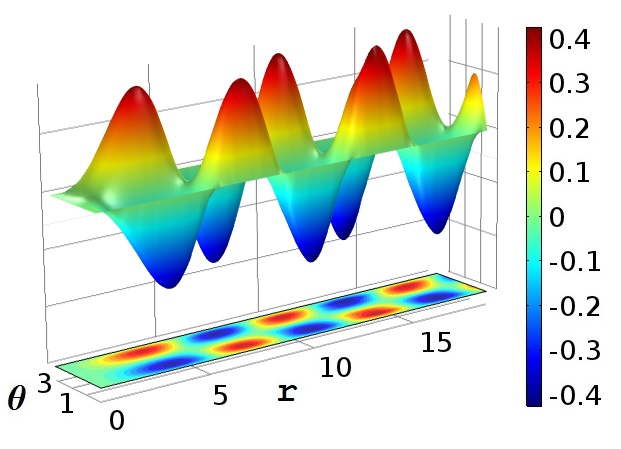}}
\hfill
\subfigure[~]{\includegraphics[width=42mm,height=30mm]{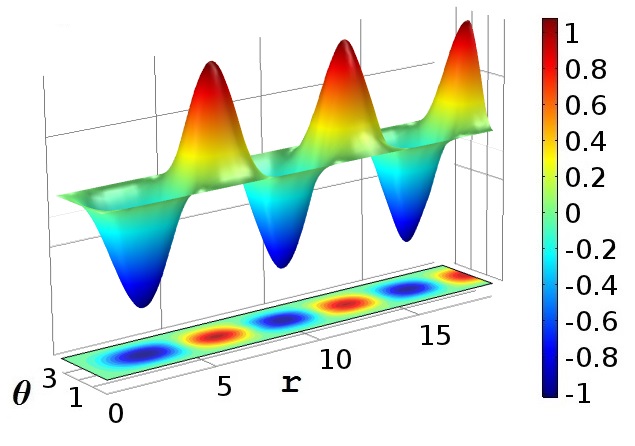}}
\hfill
\subfigure[~]{\includegraphics[width=42mm,height=30mm]{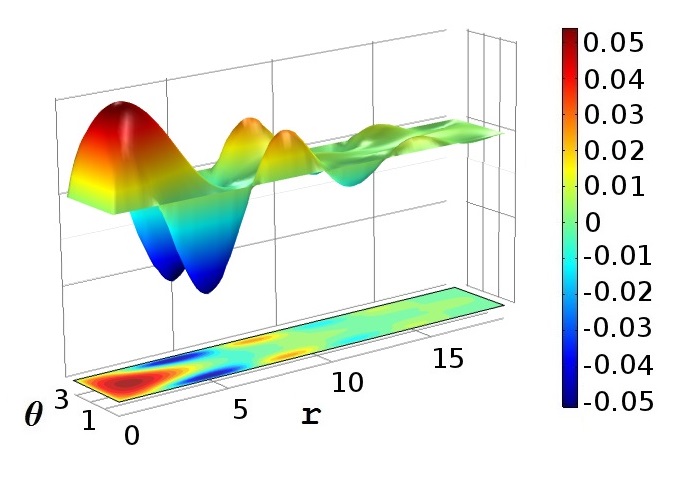}}
\hfill
%Aug9SU3ABCreg-L6C-r05C.mph
\subfigure[~]{\includegraphics[width=42mm,height=28mm]{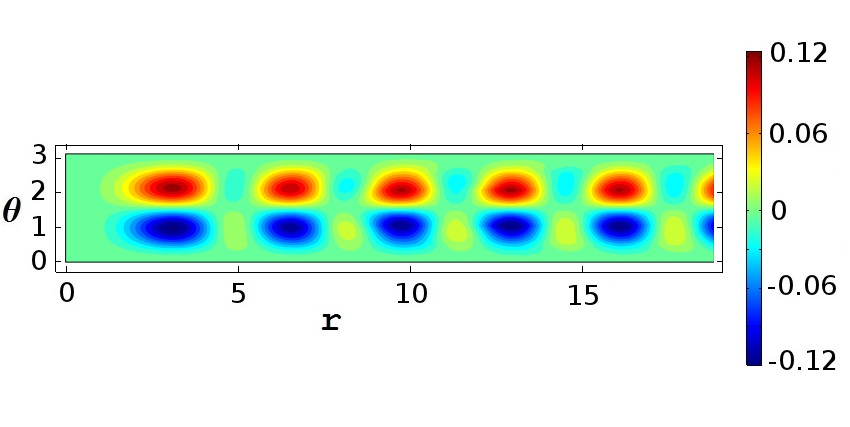}}
\hfill
%M16SU3thirdOrderextremRegKyy2-IIMMMtest.mph---su3endens
\subfigure[~]{\includegraphics[width=42mm,height=30mm]{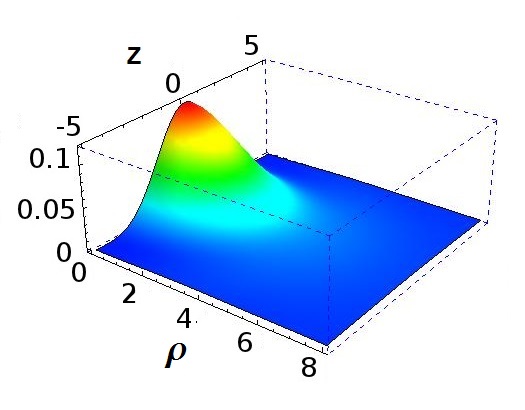}}
%M16SU3thirdOrderextremRegKyy2-IIMMM.mph
%su3ContPlapprox24
%M16SU3thirdOrderextremRegKyy2-IIMMMtest.mph
\caption[fig1]{Solution profile functions in the leading order: (a) $\tK_{1}$;
(b)  $\tK_{2}$; (c)  $\tK_{4}$; (d)   $\tK_{5}$;
(e) the time averaged radial color magnetic field $\bar B_r^3$; 
(f) the time averaged energy density $\bar{\cal E}(\rho, z)$ in the plane $(\rho,z)$ in cylindrical coordinates
($g=1,M=1$).}\label{Fig1}
\end{figure}
and it has non-vanishing time averaged radial color magnetic fields 
$\bar B^3_r=\langle F_{\theta \varphi}^3 \rangle_t
=-3/4 \tK_2\tK_4$ and $\bar B^8_r=-\bar B^3_r$, Fig. 1e,
which correspond to magnetic field distribution of a non-topological monopole-antimonopole pair 
\cite{plb2018}. The energy density of the solution is finite everywhere 
and decreases along the radial direction as $1/r^2$, Fig. 1f. 

The obtained solutions possess a nice feature, they are invariant under the Weyl permutation
group acting on $I,U,V$ sectors of $SU(2)$ subgroups. The classical Lagrangian on the space of 
field configurations satisfying the ansatz (\ref{SU3DHN},\ref{reduction1}) has an explicit Weyl symmetric form
 \bea
&&{\cal L}_{Weyl}=\sum_{p=1,2,3}\Big \{  -\dfrac{1}{6} (F_{\mu\nu}^p)^2-\dfrac{1}{2}| \pro_\mu W_\nu^p-
\pro_\nu W_\mu^p|^2 -\nn \\ 
&& \dfrac{9}{4} \Big [
 (W^{*p\mu} W^p_\mu)^2-(W^{*p\mu} W^{*p}_\mu) (W^{p\nu} W^p_\nu) \Big ]\Big \},  \label{Lphi4}
 \eea
 where index ``p'' corresponds to $I,U,V$ subgroups $SU(2)$,
 and we use complex notations for the gauge potentials \cite{SM2,flyvb,mpla2006}.  
 Note that ansatz  (\ref{SU3DHN}, \ref{reduction1}) fixes the local gauge symmetry,
 however, the Lagrangian ${\cal L}_{Weyl}$ and corresponding solutions are still invariant  
 under the action of the finite Weyl color group. While the local gauge symmetry determines 
dynamical laws by means of equations of motion, the global Weyl symmetry defines the symmetry of 
physical states and observables. In a particular, the Weyl symmetry implies an important identity,
 $A_\varphi^3 \equiv A_\varphi^8$, 
 which determines an absolute minimum of the effective potential \cite{flyvb, mpla2006},
 i.e., the vacuum energy.  
 
It is remarkable, the Weyl symmetric solutions reveal the phenomenon of 
Abelian projection.
To show this, one observes that a regular single valued solution, 
Fig. 1, is uniquely defined (up to conformal scaling transformation) by an amplitude of the Abelian field 
$K_4$ and by a choice of $\theta$-angle modes of the Fourier components $\tK_{i}(r,\theta)$, (\ref{seriesdec}). 
The Abelian projection of the solution corresponds to the limit of vanishing
off-diagonal gauge fields $K_{1,2,5}$ with a remaining Abelian potential $K_4$ which satisfies
the Maxwell type equation.

It is surprising, that the non-Abelian solutions presented in Fig. 1 and in \cite{plb2018} 
admit an Abelian dominance effect. 
A careful numeric analysis shows that a contribution of the lowest $\theta$-angle mode $\tK_4(Mr,\theta)$ to 
the total energy is $95\% \pm 1.5\%$ \cite{SM2}.
%2018Jun4SU3solABC-1L6-2extre,mph
The obtained value is very close to the known estimate of Abelian dominance 
established before only for the Wilson loop functional \cite{abeldom1,abeldom3}.
Note that an Abelian gauge potential is not itself gauge invariant.
However, the Abelian projection, applied to physical gauge invariant quantities like a 
a quantum effective action, produces gauge invariant results. 
It has been proved, that a quantum effective action with an Abelian external background field
is invariant under local gauge transformations \cite{reinhardt97}. 
 {\it We conclude, that the QCD vacuum structure 
 is determined by regular Weyl symmetric Abelian solutions which represent fixed points 
 in the configuration space under the Weyl group transformation}.
This is a direct evidence of the Abelian projection conjectured by `t Hooft \cite{thooft81}.
The existence of Weyl symmetric Abelian solutions might not be surprising,
since the Weyl group  leaves the Abelian subgroup invariant by definition.
A surprising result is that there are non-Abelian solutions which represent
fixed points under the Weyl group transformation in the whole configuration space of $SU(3)$
Lie algebra valued gauge fields. 

So far we have considered a case of axially-symmetric fileds.
In a general case, the Weyl symmetric Abelian projected QCD 
is equivalent to an Abelian gauge theory with a gauge potential $A_\mu=A_\mu^{3,8}$
corresponding to a compact structure group $U(1)$. 
A complete set of spherical wave solutions to the Maxwell equations 
is given by the vector spherical harmonics (in a temporal gauge $A_t=0$) \cite{jackson,eisenberg}
\bea
&&\vec A_{lm}^{\mathfrak{m}}=\dfrac{1}{\sqrt{l(l+1)} }\vec L j_l (k r) Y_{lm} (\theta,\varphi ) 
e^{i \omega t},\nn\\
&&\vec A_{lm}^{\mathfrak{e}}=\dfrac{-i}{\sqrt{l(l+1)} }\vec \nabla \times (\vec L j_l(k r) Y_{lm} 
(\theta,\varphi))e^{i \omega t}, \nn \\
&& \vec A_{lm}^{\mathfrak{l}}= \dfrac{1}{k} \vec \nabla j_l(kr) Y_{lm}(\theta,\varphi) e^{i \omega t},
\label{vecharmonics}
\eea
where $\omega=k c$, $\{\vec A_{lm}^{\mathfrak{m, e,l}}\}$ are eigenfunctions
of the total angular momentum operator with $J=l\geq 1$, $J_z=m$,  and the superscripts
$\mathfrak{m,e,l}$ denote magnetic, electric and longitudinal modes respectively. 
We neglect the longitudinal mode as unphysical and keep only transverse modes satisfying 
the constraints $\operatorname{Div} \vec A=0$ and $\vec E \cdot \vec B \equiv 0$.
One has still the Abelian dominance effect since the energy spectrum of general stationary
solutions is degenerated with respect to the quantum number $''m''$ due to invariance 
under the group $SO(3)$ of space rotations.

The most important step in establishing the vacuum structure 
is to prove the quantum stability of Abelian solutions (\ref{vecharmonics}) 
at microscopic level, i.e., in any small vicinity of each space-time point.
We prove this by solving a ``Schr\"{o}dinger'' 
eigenvalue equation for quantum gluon fluctuation modes $\Psi_\mu^a$ \cite{plb2018} 
\bea
&&\Ka_{\mu\nu}^{ab} \Psi_\nu^b=\lambda \Psi_\mu^a, \label{schr3} \nn \\
&&\Ka_{\mu\nu}^{ab}=-
\delta^{ab} \delta_{\mu\nu} \pro^2_t
            -\delta_{\mu\nu}({\Da}_\rho {\Da}^\rho)^{ab} -2 f^{acb}{\mathcal
	    F}_{\mu\nu}^c, \quad \label{Koper}
\eea
where the operator $\Ka_{\mu\nu}^{ab} $ corresponds to one-loop gluon contribution 
to the effective action \cite{plb2018}, and $\Da_\mu, \Fa_{\mu\nu}$ 
are defined by means of background Abelian solution (\ref{vecharmonics}).  
The existence of a negative eigenvalue $\lambda$
would indicate to the presence of an unstable mode which destabilizes the vacuum.
We solve numerically the eigenvalue equation (\ref{schr3}) with
an axially-symmetric background Abelian field, $m=0$,
for various values of the parameters $(l,M)$ and radial size $L$
of the numeric space domain.  
Our numeric analysis confirms the 
positiveness of the spectrum of the operator $K_{\mu\nu}^{ab}$
for parameter values 
$l\leq 4$ and $\nu_{11}\leq M \leq 10$, $\nu_{11}\leq L \leq 200$.
Typical time dependent fluctuation modes 
corresponding to the lowest eigenvalue are shown in Fig. 2 
(details of calculation are enclosed in \cite{SM2}). 
It is somewhat unexpected, that standing wave solutions 
of electric type are stable as well \cite{SM2}, contrary to a 
case of constant chromoelectric field \cite{schan}. 
This proves that transverse Abelian solutions (\ref{vecharmonics})
can serve as structural elements of a stable vacuum.
%Feb20AbelMagnStabeqs.nb
%Feb20Feb32018stabeqsCOMSOL.rtf
%Feb20Feb8SU2AbelStabL6M1A1reg.mph
\begin{figure}[h!]
 \centering
\subfigure[~]{\includegraphics[width=42mm,height=30mm]{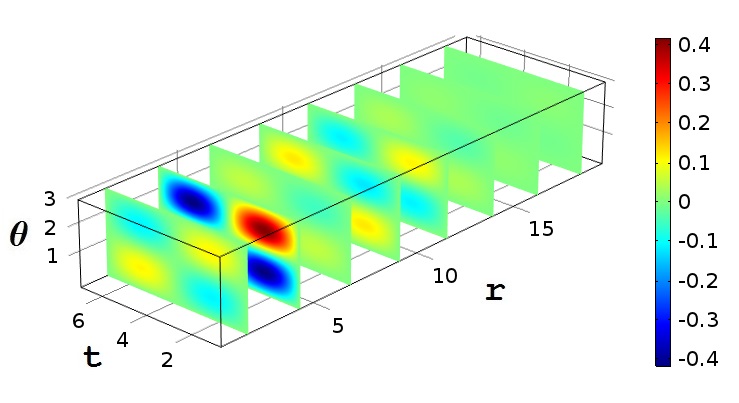}}
\hfill
\subfigure[~]{\includegraphics[width=42mm,height=30mm]{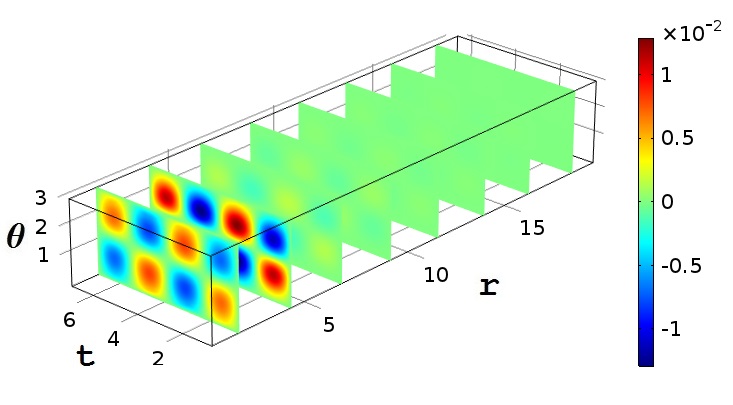}}
\caption[fig2]{Gluon fluctuation modes with the lowest eigenvalue  $\lambda=0.0292$
for the vacuum magnetic field with ($l=1, m=0)$: 
(a) $\Psi_3^2$; (b) $\Psi_4^1$ ($g=1,M=1$).
}\label{Fig2}
\end{figure}
%In particular, the general basis in the space of regular electric type 
%solutions corresponding to the lowest energy, $l=1$, is the following
%Feb15AbelMaxwMagnElectypesCHCK.nb
%Feb15AbelMaxwMagnElectypes.nb, zamena p1->c1,p4->c2,c43->c3

We quantize the Abelian solutions (\ref{vecharmonics}) in a finite space 
region constrained by a sphere of radius $a_0$ corresponding to an effective
glueball size.
It is suitable to introduce dimensionless units $\tilde M=M a_0, x=r/a_0, \tau=t/a_0$.
To find proper boundary conditions we require that the Pointing vector 
$\vec {\mathbf S}=\vec {\mathbf E} \times \vec {\mathbf B}$  
vanishes on the sphere. This implies two possible types of boundary
conditions
\bea
(I): &\vec A_{lm}^{\mathfrak{m,e}} (\tilde M x)|_{x=1}=0, ~ \tilde M_{nl}=\mu_{nl},  \nn \\
(II): &\pro_r(r \vec A_{lm}^{\mathfrak{m,e}} (\tilde M x))|_{x=1}=0, ~ \tilde M_{nl}=\nu_{nl},  \label{munu}
\eea
where $\mu_{nl}$ are nodes of the Bessel function $j_l(r)$, and $\nu_{nl}$ denote zeros of  
the function $\pro_r (r j_l(r))$, the integers $(n\geq 1 ,l\geq 1)$ are the main and 
orbital quantum numbers. 
Our approach resembles the old MIT bag models \cite{MIT1,MIT2,MIT3} where the energy spectrum 
of free glueballs was obtained for the first time \cite{jaffe1976,johnson1981,jaffe1986}. 
However, there are principal differences between our formalism and bag models: (i)  
we apply the Weyl symmetric Abelian projection which leads directly to physical quantum 
states; (ii) in bag models one has to impose an additional constraint
$ n_\mu F^{\mu\nu}=0$, ($n_\mu=(0, \vec n \equiv \vec r/r)$) \cite{MIT3} which 
prevents the color charge to pass through the bag surface.

We choose the following normalization condition for the
vector harmonics $\vec A_{\tM lm}^{\mathfrak{m, e}}$ 
\bea
\dfrac{1}{4\pi}\int_0^1 dx \int d\theta d\varphi  x^2 \sin\theta(\{\vec A_{\tM_{nl} lm}^{\mathfrak{m, e}})^2
=\dfrac{1}{\tM_{nl}}. \label{normcond}
\eea
The standard canonical quantization results in the following Hamiltonian
expressed in terms of the creation and annihilation operators   $c^{\pm}_{nl}$
\bea
H= \dfrac{1}{2} \sum_{n,l,m} \tilde M_{nl} ( c^+_{nlm} c^-_{nlm}+ c^-_{nlm} c^+_{nlm}). \label{Hamvac}
\eea
One particle states $\{c^+_{nlm} |0\rangle\}$ describe primary free Abelian glueballs.
%Mar21QuantFin1CHCKfin.nb
The knowledge of explicit solutions for the vector potential (\ref{vecharmonics})
allows to estimate densities of various vacuum gluon condensates.
Averaged over the time and polar angle vacuum gluon condensates
  $\alpha_s\langle (F_{\mu\nu})^2 \rangle$ corresponding to magnetic modes 
  $\vec A_{\nu_{11} 11}^{\mathfrak{m}}$ and  $\vec A_{\mu_{11} 11}^{\mathfrak{m}}$
    are depicted in Fig. 3 ($\alpha_s=0.5$ \cite{PDG}).
 \begin{figure}[h!]
%Oct10Aug24Mar21QuantFin1CHCKfin.nb
{\includegraphics[width=68mm,height=42mm]{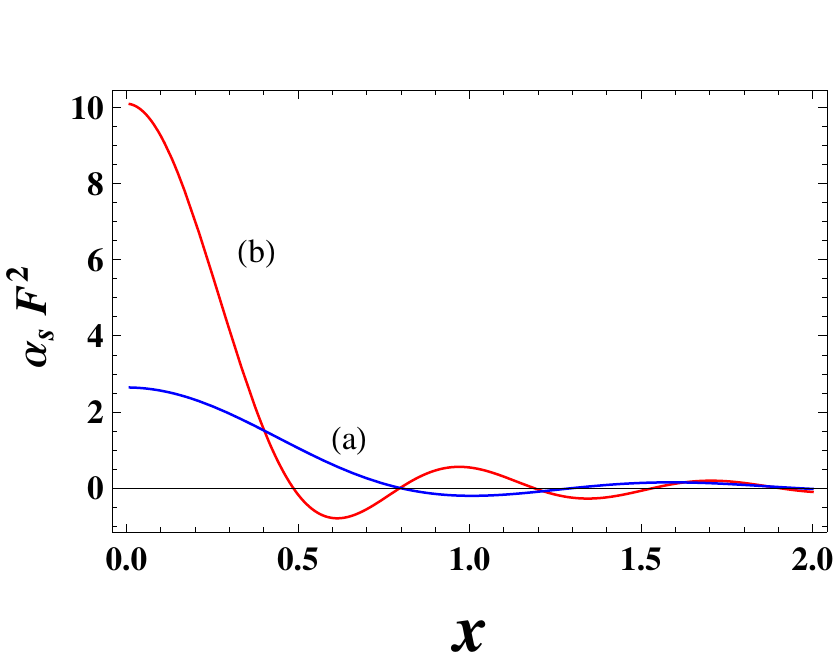}}
\caption[fig3]{Radial densities of the magnetic vacuum gluon condensates $\alpha_s\langle \overline{F^2} \rangle$ 
corresponding to modes $\nu_{11}=2.74...$, (a), and $\mu_{11}=4.49...$, (b); (n=l=1, m=0).}\label{Fig3}
\end{figure}
%Mar21QuantFin1CHCKfin.nb
The oscillating behavior of the vacuum gluon condensate density was obtained before within 
the instanton approach to QCD \cite{dorokhov1997,dorokhov2000}.
Integrating the gluon condensate density over the interval $(0\leq x \leq 1)$ 
one can fit the value of the vacuum gluon condensate, $\alpha_s\langle (F_{\mu\nu}^a)^2 \rangle=(540MeV)^4$,
by fixing the effective glueball size parameter, $a_0=0.376 Fm$. 
%Aug19LocSol-highnodes.nb
With this, the energy spectrum of free Abelian glueballs $J^{PC}$ is $E_{nl}=197 \tilde M_{nl}/a_0$ MeV
($J=l$, $P=(-1)^{J+1}$ for 
magnetic glueballs and $P=(-1)^J$ for electric glueballs, $C=-1$ \cite{jackson,eisenberg}).
%Jun20GlSpectr.nb
%Mar21QuantFin1CHCKfin.nb
Two lightest glueballs,  $(1)^{+-}, (1)^{--}$ of magnetic and electric types respectively, 
have the same energy $E_{11}=1.435$ GeV corresponding to antinode $\nu_{11}$.

Certainly, the primary free glueballs are not observable quantities since we have not taken into account 
their interaction to vacuum gluon condensate. 
To estimate the effect of such interaction we use a method applied in study of the
anomalous magnetic moment of a photon propagating in the external magnetic field 
\cite{magnphoton1, magnphoton2}. In QCD the role of an external magnetic field is 
played by the magnetic vacuum gluon condensate.  
Consider one-loop $SU(3)$ effective Lagrangian ${\cal L}^{1-l}$ in constant field approximation 
\cite{savv, flyvb}
\bea
{\cal L}^{1-l} =-\dfrac{1}{4} \tF^2-k_0 g^2 \tF^2 \Big (\log (\dfrac{g^2 \tF^2}{\Lambda_{QCD}^4})-c_0 \Big ), 
\label{Leff}
\eea
where $\tF_{\mu\nu}$  is an external Weyl symmetric Abelian color magnetic field, 
the factor $k_0=\dfrac{11}{64 \pi^2}$ is proportional to 1-loop beta function, 
and $c_0$ is a renormalization constant. A corresponding one-loop effective potential implies
a value  $\simeq (70MeV)^4$ for the vacuum gluon condensate, which is not reliable.
However, the most important property of the effective Lagrangian is the presence of 
essentially non-perturbative structure which leads to appearance of a non-trivial minimum
of the effective potential. Such a structure allows to derive an effective Lagrangian for physical
glueballs with setting $k_0, c_0$ as free phenomenological parameters.
% Jul23BndSt-short.nb
Following the method in \cite{magnphoton1, magnphoton2}
we split the external field $\tF_{\mu\nu}$ into two parts
\bea
\tF_{\mu\nu}=B_{\mu\nu} + F_{\mu\nu}, 
\eea
where $B_{\mu\nu}=\pro_\mu B_\nu-\pro_\nu B_\mu$ describes the magnetic vacuum gluon condensate, and 
$F_{\mu\nu}=\pro_\mu A_\nu-\pro_\nu A_\mu$ contains the Weyl symmetric Abelian gauge potential
$A_\mu$ describing the physical glueball.
In approximation of slowly varied vacuum condensate we decompose the Lagrangian ${\cal L}^{1-l} $
around the vacuum field $B_{\mu\nu}$ 
%Aug30Jul23BndSt.nb
corresponding to a non-trivial minimum of the effective potential at the value of vacuum
gluon condensate
\bea
g^2 B^2 =\Lambda_{QCD} \exp \big(c_0-1-\dfrac{1}{2k_0g^2}\big).\label{npcond}
\eea
With this one obtains an effective Lagrangian for physical Abelian glueballs in quadratic approximation
\bea
{\cal L}_{eff}^{(2)}[A]=-2 k_0 g^2 \dfrac{(B^{\mu\nu} F_{\mu\nu})^2}{B^2} \equiv
-\kappa (B^{\mu\nu} F_{\mu\nu})^2 , \label{LagrBf}
\eea
where we neglect a term corresponding to an absolute
value of the vacuum energy, and $\kappa$ is an effective coupling constant.
The effective Lagrangian ${\cal L}_{eff}^{(2)}[A]$ is strikingly different from a corresponding effective Lagrangian 
in QED \cite{magnphoton1, magnphoton2}. Namely, the expression (\ref{LagrBf}) does not include the 
classical kinetic term $-1/4 F_{\mu\nu}^2$ which has disappeared due to non-perturbative origin of the 
vacuum gluon condensate (\ref{npcond}). Note that result (\ref{LagrBf}) is model independent,
and it can be obtained from a class of Lagrangian functions which admit series expansion around
a non-trivial vacuum.

In a case of the lightest magnetic glueball the vacuum gluon condensate 
is described by the vector harmonic $\vec A_{l=1,m=0}^{\mathfrak{m}}$,
 $\omega=\nu_{11}$, with one non-zero component
\bea 
B_\varphi(x,\theta,\tau)=N_{11}x j_1(\nu_{11} x)\sin^2 \theta \sin(\nu_{11}\tau),
\eea
where $N_{11}$ is a renormalization constant.
The lightest magnetic glueball is described by the gauge potential 
$A_\varphi(x,\theta,\tau)$ which assumed to be time-coherent 
to the vacuum condensate field, $A_\varphi(x,\theta,\tau)=a(x,\theta) \sin(\nu_{11} \tau +\phi_0)$,
with a constant phase shift $\phi_0$. A time-averaged effective Lagrangian ${\cal L}_{eff}^{(2)}[A]$ 
leads to Euler equation for the field $a(x,\theta)$ \cite{SM2}.  
The equation is separable and implies an ordinary differential equation 
for a spherically symmetric function
$a(x,\theta)\equiv f(x)$ 
\bea
&& r^2 b'^2 f'' -2b'(b+x b'-x^2 b'')f' - \nn \\
&&\nu_{11}^2 \Big (\nu_{11}^2 x^2 b^2-\xi \nu_{11}^2 x^3 j_2(\nu_{11} x)  b' 2\Big ) f=0, \label{eqf}
\eea
where $b(x)=\nu_{11} x j_1(\nu_{11}x)$,  $\xi\equiv\cos(2\phi_0)/(2+\cos(2\phi_0))$.
The coefficient functions in front of the first and second derivative terms in (\ref{eqf}) 
vanish at $x_0=1$ (or $r_0=a_0)$. This implies that a regular solution is localized 
inside a finite interval $0\leq x \leq 1$, Fig. 4a. 
%Aug22Aug15LocSol-3impr.nb
%Aug16LocSol-3impr.nb,Aug15LocSol-3impr.nb
%Aug17LocSol-1ksid-1reg.mph,Aug16LocSol-1ksid-1reg.mph
 \begin{figure}[h!]
\centering
{\includegraphics[width=70mm,height=44mm]{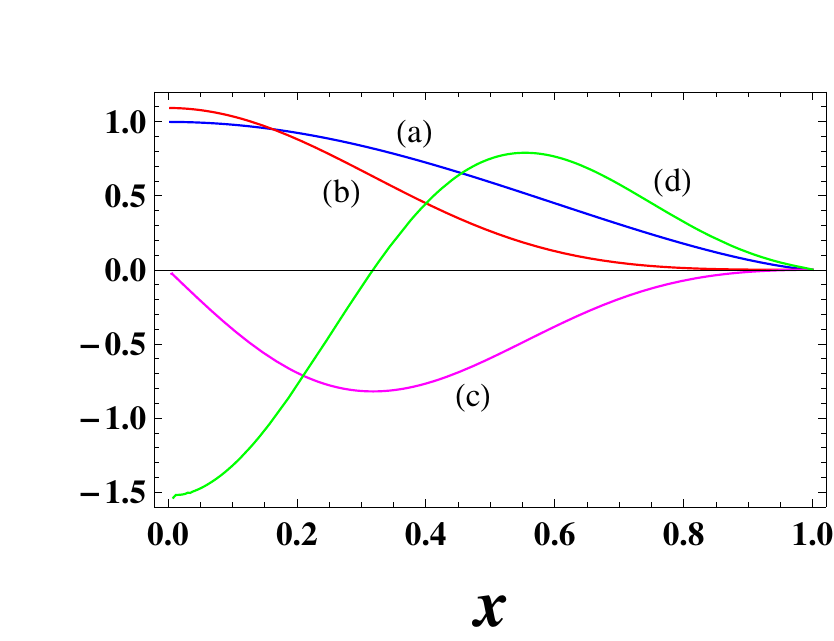}}
\caption[fig4]{(a) Solution $f(x)$ with a unit initial amplitude; (b) a corresponding radial energy density 
$\bar{\cal E}/4 \pi \kappa$; 
(c) the first derivative of the energy density; (d) the second derivative of the energy density; $\phi_0=\pm \pi/2$.      
}\label{Fig4}
\end{figure}
The solution has a removable singularity at $x_0=1$. To verify that solution is physical we 
check the properties of the energy density averaged over the time and polar angle
\bea
\bar{\cal E}&=&\dfrac{\pi \kappa}{4 x^2 } \Big ((2+\cos(2 \phi_0)) b^2 f^2 +2 \cos(2 \phi_0)  b b' f f' \nn \\
&& +(2+\cos(2 \phi_0)) b'^2 f'^2 \Big ).
\eea
One can verify that the energy density and its first and second derivatives 
are continious functions at $x_0=1$, Fig. 4bcd.
The solution with the phase shift value $\phi_0=\pm\pi/2$ has a minimal energy,
and the corresponding effective Lagrangian vanishes identically as for the photon plane wave. 
So the obtained solution for the physical glueball
is stable, it does not collapse and has the energy upper bound $1.435$ GeV
provided by the energy of the lightest free glueball.

In QED the photon anomalous magnetic moment leads to changing 
the total angular momentum \cite{magnphoton2}. A similar effect in QCD provides
the total angular momentum $J=0$ for the physical glueball. 
Quantum numbers of the glueball interacting to vacuum magnetic gluon condensate
are defined in the same way as for two photon system \cite{landau2photon,yang2photon,minkowski2photon}
and lead to two lightest magnetic glueballs $0^{++}, 0^{-+}$ which might be identified with the
expected lightest two-gluon glueballs \cite{kochelev2009,ochs2013}. 
Note that the lightest free electric glueball $1^{+-}$ does not interact to vacuum magnetic condensate
due to the structure of the effective Lagrangian ${\cal L}_{eff}^{(2)}[A]$. So that it is fictious 
in agreement with the Yang theorem \cite{ochs2013, yang2photon}.

In conclusion, the proposed microscopic description of a pure QCD vacuum
reveals the Weyl symmetric Abelian projection and Abelian dominance phenomenon.
This allows to describe a full pure glueball spectrum endowed with a fine structure 
generated by off-diagonal gluons. A detailed structure of the pure glueball spectrum
and generalization to QCD with quarks will be considered in a separate paper.

 \acknowledgments

Authors thank N.I. Kochelev, A. Pimikov, J. Evslin, A.B. Voitkiv, A. Kotikov for 
useful discussions. This work is supported by the National Natural Science Foundation of China
(Grant No. 11575254).

\end{document}